\documentclass[]{aastex631}
\usepackage{textcomp}
\usepackage{gensymb}
\usepackage{amsmath}
\usepackage{url}
\usepackage{color}

\shorttitle{TRGB zero-point calibration with \textit{Gaia} EDR3}
\shortauthors{Li et al.}

\graphicspath{{./}{figures/}}
\DeclareUnicodeCharacter{2212}{-}
\begin{document}

\title{A Maximum Likelihood Calibration of the Tip of the Red Giant Branch Luminosity from High Latitude Field Giants using Gaia Early Data Release 3 Parallaxes}

\correspondingauthor{Siyang Li}
\email{sli185@jhu.edu}

\author[0000-0002-8623-1082]{Siyang Li}
\affiliation{Department of Physics and Astronomy, Johns Hopkins University, Baltimore, MD 21218, USA}

\author{Stefano Casertano}
\affiliation{Space Telescope Science Institute, 3700 San Martin Drive, Baltimore, MD 21218, USA}

\author[0000-0002-6124-1196]{Adam G. Riess}
\affiliation{Department of Physics and Astronomy, Johns Hopkins University, Baltimore, MD 21218, USA}
\affiliation{Space Telescope Science Institute, 3700 San Martin Drive, Baltimore, MD 21218, USA}

\begin{abstract}
The calibration of the tip of the red giant branch (TRGB) in the $I$-band
has a direct role in determinations of the Hubble constant, a subject of recent interest due to the discrepancy between direct and indirect estimates of its value.  We present a maximum likelihood (ML) method designed to obtain an independent calibration of the brightness of TRGB  using \textit{Gaia} parallaxes from the Early Data Release 3 (EDR3) of Milky Way field Giants at high Galactic latitude.  We adopt simple parameterizations for the Milky Way stellar luminosity function and density law and and optimize the likelihood of the observed sample as a function of those parameters.  Using parameters to partially constrain the luminosity function from other galaxies similar to the Milky Way for which high quality TRGB data are available,  we find values of the TRGB magnitude of $ M_I^{TRGB} = -3.91 \pm 0.05 $ (stat) $ \pm 0.09 $ (sys)~mag, where the systematic uncertainty covers the range of shape parameters found in our Milky Way sample and in reference galaxies. While APASS Data Release 9 all-sky photometry is insufficient to provide a reliable constraint on the shape of the Milky Way luminosity function, we estimate that the photometry from \textit{Gaia} Data Release 3 (mid-2022) will allow better constraints on the shape, and lower statistical uncertainties on the tip by a factor of 3.  With expected releases of improved parallax measurements from $Gaia$, the method of calibrating the TRGB luminosity from field Giants is expected to reach $\sim$ 0.01~mag uncertainty, which is an important step toward a precise TRGB-based determination of the Hubble constant.

\end{abstract}


\section{Introduction} \label{sec:intro}

The tip of the red giant branch (TRGB) is a luminous and widely used standard candle that is largely insensitive to stellar mass and metallicity near the $I$-band \citep{Iben_1983ARA&A..21..271I, Ferraro_2000AJ....119.1282F, Salaris_2002PASP..114..375S, Rizzi_2007ApJ...661..815R, Bellazini_2008MmSAI..79..440B} and has been used to measure extragalactic distances \citep{Sandage_1971swng.conf..271S, Graham_1982ApJ...252..474G, Mendez_2002AJ....124..213M, Salaris_2005MNRAS.357..669S, Makarov_2006AJ....132.2729M, Anand_EDD_2021AJ....162...80A} and more recently in distance ladders to measure the Hubble constant, $H_0$ \citep{,Jang_2017ApJ...835...28J,Freedman_2019ApJ...882...34F}.  Interest in the Hubble constant has grown due to the 4 - 6 $\sigma$ discrepancy between direct, local measurements of $H_0$ and its value predicted using the cosmic microwave background and the cosmological model \citep{Planck_2020A&A...641A...6P, Riess:2022}. Measurements of $H_0$ which include the use of TRGB vary from 70--77 km s$^{-1}$ Mpc$^{-1}$ depend on the calibration of the TRGB and other components of the distance ladder \citep{Freedman_Tensions_2021ApJ...919...16F, Anand_EDDvsCCHP_2021arXiv210800007A, Soltis_2021ApJ...908L...5S, 2022arXiv220107801J,2022arXiv220304241D,blakeslee:2021}.   The TRGB is more challenging to calibrate than many other standard candles because it is a feature in the luminosity function rather than a class of star, so that its value cannot be estimated from parallaxes of the nearest examples.  Rather efforts have been limited to a calibration derived from a co-located distribution of stars. Here we present a method to obtain a calibration using an ensemble of all-sky parallax determinations from $Gaia$ EDR3 of field Giants.

Theoretically, the nature of the TRGB feature of the luminosity function of stars is well understood. Red giant stars with masses less than $\sim$ 2 $M_{\odot}$ that do not have adequate pressure to immediately begin fusing helium at the end of hydrogen fusion in their cores continue to fuse hydrogen in a shell around an inert, isothermal helium core supported by electron degeneracy pressure. As the core grows from the deposition of helium from the hydrogen burning shell, the radius of the core remains constant, resulting in an increase in density and temperature. Once the temperature of the core reaches a critical temperature, helium burning is triggered. Because pressure in electron degeneracy does not vary with temperature, although the temperature in the core increases, there is no immediate expansion. This results in a helium burning runaway leading to the Helium Flash \citep{Iben_1983ARA&A..21..271I}. During the Helium Flash, increasing temperature begins to decrease degeneracy which allows for the core to expand, and the star rapidly increases in brightness before quickly fading. This event results in a discontinuity in the luminosity function (LF) on the giant branch, which is often identified as a `tip' of the red giant branch either in the color-magnitude diagram or in the luminosity function. After the Helium Flash, the core is no longer degenerate and the star transitions onto the zero age horizontal branch. 

Early efforts to identify the TRGB were primarily based on visual inspection of color-magnitude diagrams in M31 and M33 \citep[see, for instance, ][]{Mould_1986ApJ...305..591M}. The first published effort at establishing a reproducible and quantitative measurement of the TRGB was made by \cite{Lee_1993ApJ...417..553L} who applied a zero-sum Sobel kernel of the form [-2, 0, 2] to the LF of red giant populations in 10 galaxies. This method tries to detect an edge or transition in the luminosity function. Since then, the Sobel filter approach has been extended and improved upon in several ways. For instance, \cite{Sakai_1996ApJ...461..713S} used a continuous probability distribution to smooth the LF according to photometric error and an extended Sobel filter of [-2, -1, 0, 1, 2]. Similar to \cite{Sakai_1996ApJ...461..713S}, \cite{Freedman_2019ApJ...882...34F} and \cite{Jang_2018ApJ...852...60J} also smooth their LF, though with a nonparametric Gaussian-windowed, Locally Weighted Scatterplot Smoothing  \citep[GLOESS;][]{Hatt_2017ApJ...845..146H}, and use a [-1, 0, 1] kernel instead of an extended, five unit kernel. A more thorough review of the use of Sobel filters to identify the TRGB can be found in \cite{Beaton_2018SSRv..214..113B}.

The Sobel filter approach is typically restricted to the case in which all stars can be assumed to be at the same distance, and the measured apparent magnitude maps directly to absolute magnitude (with an offset to be determined).  We refer to this as the `one-dimensional' case, in which only the probability density in magnitude space is considered.  When applied to populations where both magnitudes and distances vary, such as for Milky Way field stars, the Sobel method becomes unreliable due to the sparsity of stars in each two-dimensional bin and the complex effect of parallax uncertainties.  It is also more sensitive to noise and to additional discontinuities in the luminosity function.   Due in part to these additional complexities, Milky Way field stars have not yet been successfully used to calibrate the TRGB zero-point \citep[but see][]{Mould_2019PASA...36....1M}.

An alternative method based on maximum likelihood (ML) has also been applied to measure the TRGB in the one-dimensional case.  \cite{Mendez_2002AJ....124..213M} use ML to locate the TRGB in various galaxies and adopt a power law LF following \cite{Zoccali_2000A&A...358..943Z} instead of the binned LF used in the Sobel filter approach. The authors find close agreement between the two methods, with ML yielding lower uncertainties; this suggests that the ML approach can yield more precise estimates of the TRGB than the Sobel filter method under appropriate conditions.  \cite{Makarov_2006AJ....132.2729M} apply a similar ML approach to locating the TRGB in a sample of artificial galaxies and \textit{Hubble Space Telescope} ($HST$) WFC2 and ACS observations of 13 dwarf galaxies; they modify the method used by \cite{Mendez_2002AJ....124..213M} by applying a photometric error function from artificial star tests to mitigate the effects of photometric errors, blending, and incompleteness in stellar populations \citep{Dolphin_2002MNRAS.332...91D}, as well as implementing a different power law distribution for the luminosity function. The authors note that ML is less sensitive to undersampling bias compared to the Sobel filter method.  The method is used to measure distances to over 300 galaxies by \cite{Anand_EDD_2021AJ....162...80A}.   \cite{McQuinn_2014ApJ...785....3M} and \cite{McQuinn_2021ApJ...918...23M} also use ML to find the TRGB of dwarf galaxies for the Survey of H I in Extremely Low-mass Dwarfs (SHIELD), adopting an approach similar to \cite{Makarov_2006AJ....132.2729M}.

To date, the ML approach has only been applied in the one-dimensional case. However, it is possible to extend the ML approach to the two-dimensional case, in which the joint distribution of the target stars in distance and absolute magnitude is considered.  For Milky Way field stars, absolute magnitude and distance map to, and are constrained by, observed apparent magnitude and parallax, making it possible to formulate a model and optimize its parameters given the observed sample.  The formal basis of the method is given in Section~\ref{ss:ML_details}; here we note that any selections based on observed quantities can be properly accounted for by ensuring that the same selections are applied to the likelihood calculation, and especially to its normalization.  With this method, we can use Milky Way field stars to calibrate the TRGB zero-point. While \cite{Mould_2019PASA...36....1M} have attempted to use Milky Way fields stars to calibrate the TRGB zero-point with a chi-squared approach, it is difficult to evaluate its effectiveness due to the lack of uncertainty on their quoted TRGB value of $\sim -4.0$~mag.

Recent advances in the precision and coverage of all-sky surveys have made this two-dimensional ML approach viable for a sample of Milky Way stars.  $Gaia$ is a European Space Agency project that launched in 2013 and has obtained astrometry and photometry of over one billion stars in the Milky Way. Their most recent data release, Early Data Release 3 \citep[EDR3;][]{Gaia_2021A&A...649A...1G}, provides the most precise parallaxes to Milky Way stars to date. The American Association of Variable Star Observers (AAVSO) began the AAVSO Photometric All-Sky Survey (APASS)\footnote{\url{https://www.aavso.org/apass}} in 2010 and uses two 20 cm astrographs at each observatory at the Dark Ridge Observatory in New Mexico and the Cerro Tololo Inter-American Observatory (CTIO) in Chile. Observations are performed in five bandpasses (Johnson $ BV $ and Sloan $g' r' i'$) and fill in the gaps between the shallower Tycho2 and fainter surveys such as the Sloan Digital Sky Survey (SDSS) and Panoramic Survey Telescope and Rapid Response System (PanSTARRS) \citep{Henden_2018AAS...23222306H}. APASS provides photometry for a large sample of Milky Way fields stars that does not saturate in the range of magnitudes relevant for our TRGB selection.

In this study, we present a calibration of the TRGB zero-point based on the application of the two-dimensional ML approach to a sample of high-latitude Milky Way field stars. 
In Section \ref{sec:DataSelection}, we describe the selection criteria and cuts used to obtain our sample of Milky Way field stars. In Section \ref{sec:ML}, we describe the two-dimensional ML approach we use to identify the TRGB zero-point and the ML optimization constraints in Section \ref{sec:Optimization}. Finally, in Section \ref{sec:Disc} we discuss the assumption made in our estimation and potential for improvement with future $Gaia$ data releases.  

\section{Data Selection}
\label{sec:DataSelection}

We begin by selecting targets from $Gaia$ EDR3\footnote{Retrieved at \url{https://gea.esac.esa.int/archive/}} relevant to the TRGB on the basis of their $Gaia$ magnitude $ G $ , $Gaia$ color $ G_{BP} - G_{RP} $, and Galactic latitude $ b $: $ G \leq 13 $~mag, $G_{BP} - G_{RP} \geq 1.3 $~mag, and $ | b | \geq 36^\circ $, respectively.  The Astronomy Database Query Language (ADQL) query used for this selection is

\begin{verbatim}
SELECT * FROM gaiaedr3.gaia_source
WHERE (b >= 36 OR b <= -36)
AND phot_g_mean_mag <= 13
AND bp_rp >= 1.3
\end{verbatim}

\begin{figure}[ht!]
\plotone{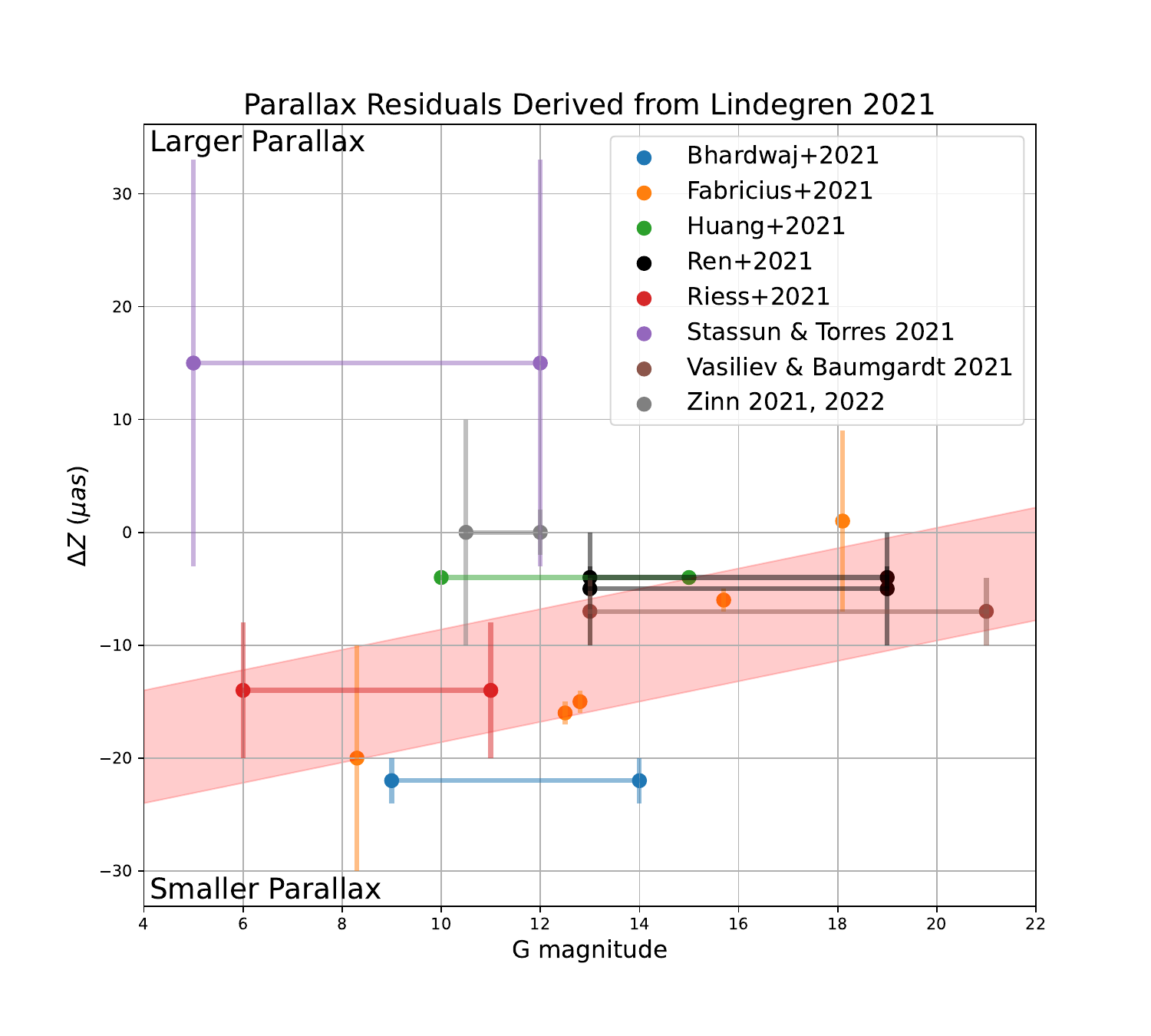}
\caption 
{$Gaia$ parallax offset residuals measured externally by sources compiled by \cite{Lindegren_2021_talk}. The fit for the $ 10 < G < 13 $~mag range is represented by the red shaded region where $\pm$ 5 $\mu$as to the fit has been added for clarity. The offsets in the plot are from \cite{Bhardwaj_f2021ApJ...909..200B, Fabricius_2021A&A...649A...5F, Huang_2021ApJ...910L...5H, Ren_2021AJ....161..176R, Riess_2021ApJ...908L...6R, Stassun_2021ApJ...907L..33S, Vasiliev_2021MNRAS.505.5978V, Zinn_2021AJ....161..214Z}.}
\label{fig:PaxOffset}
\end{figure}

These three cuts remove stars that are far from the TRGB in magnitude and color and located near the Galactic plane where crowding can bias our results. The high latitude selection limits Milky Way extinction to a mean of $A_I$ = 0.063~mag as determined following \cite{Schlafly_2011ApJ...737..103S}.
This initial set of cuts leaves 125,424 stars, many of which are nearby main sequence stars or on the lower giant branch.  We then crossmatch these stars with APASS Data Release 9 using the Strasbourg astronomical Data Center (CDS) crossmatch service \citep{Boch_2012ASPC..461..291B, Pineau_2020ASPC..522..125P} and a fixed crossmatch radius of 5 arcsec. We remove targets that have an $i'$ or $r'$ magnitude error of 0 mag in the crossmatch results and do not use APASS Data Release 10 because AAVSO describes it as missing the northern polar cap and having photometric errors larger than those in Data Release 9\footnote{As of writing this manuscript, this description can be found at \url{https://www.aavso.org/apass}}. This second selection leaves 90,446 stars. We then apply parallax zero-point offset corrections following \cite{Lindegren_2021AA...649A...4L} and exclude stars that do not have an available correction, which leaves 89,655 stars. The sample of stars used in this study mostly fall in the $ 10 < G < 13 $~mag, in which the parallax offset has been determined by \cite{Lindegren_2021AA...649A...4L}.  However, this range is brighter than most of the sources used to calibrate the offset (e.g., quasars) and several studies have reported parallax residuals with respect to the \cite{Lindegren_2021AA...649A...4L} correction; Fig. \ref{fig:PaxOffset} shows a selection of these results as compiled by \cite{Lindegren_2021_talk}.  The residual offset $\delta \varpi $ in the region $ 10 < G < 13 $~mag is consistently negative (true parallaxes are smaller) and is well approximated by
\begin{equation}
    \delta \varpi \sim -10 - 0.9\cdot (14-G) \, \mu{\rm as} \pm 5 \mu{\rm as}
    \label{eq:delta_offset}
\end{equation}  We add this residual from the EDR3 parallaxes after applying the prescribed \cite{Lindegren_2021AA...649A...4L} correction.

The magnitude of the TRGB is usually reported in the Johnson-Cousins $ I $ band.  To match this band, we convert Sloan $i'r'$ magnitudes from APASS to Johnson-Cousins $I$ magnitudes using photometric conversions provided by SDSS \footnote{\url{http://classic.sdss.org/dr5/algorithms/sdssUBVRITransform.html\#Lupton2005}}. We first convert $i'r'$ to $ir$ using

\begin{equation}
    i = i' + 0.41((r'-i') - 0.21); \qquad \hbox{$r = r' + 0.35((r'-i') - 0.21)$}
\end{equation}

\noindent then convert $i$ to I using

\begin{equation}
    I = r - 1.2444(r - i) - 0.382
\end{equation}

\noindent We add the filter transformation uncertainty of 0.0078 mag to the propagated uncertainties. We then apply extinction corrections from the \cite{Schlafly_2011ApJ...737..103S} dust maps using the NASA/IPAC Galactic Dust Reddening and Extinction tool. We find 36 stars that do not have corresponding extinction data from this tool and exclude these stars from our sample. Errors for the extinction corrections are estimated using the standard deviation of the pixel values and are on average $2 \pm 5$~mmag or $\sim 5 \%$ of the extinction. We then restrict the sample to stars having an apparent $ I $-band magnitude between 8.6~mag and 11.7~mag to avoid incompleteness bias; the sample is expected to be essentially complete within this range, and the bright cut at $ I = 8.6 $~mag avoids saturation issues with APASS.  We exclude stars with parallax and magnitude errors larger than $ 0.5 $~mag and mas; because of the large uncertainties, those stars contribute little to the likelihood calculation, and can create numerical inaccuracies.  Adopting a more conservative cut for magnitude error of 0.2~mag would decrease the final number of stars to 3,806 and shift our final TRGB measurement by 1~mmag in the brighter direction.

The final cut we implement is in measured parallax.  The goal of this selection is to include as many stars close to the TRGB as possible, without diluting the sample excessively.  For this purpose, we exclude stars whose parallax corresponds to a {\it nominal} absolute magnitude outside the range $ [-5, -3] $~mag (the blue and red lines in Fig.~\ref{fig:absrange}).  This cut is applied in {\it observed} parallax space; to ensure that no bias in the likelihood calculation is introduced, the same selection must be and is applied in the computation of the 
normalization of the likelihood (see Eq.~\ref{eq:normalization_integral}).

\begin{figure}[ht!]
\plotone{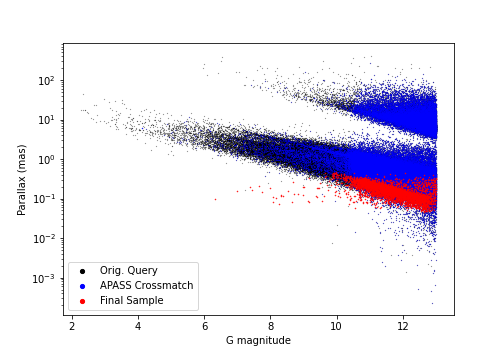}
\caption 
{Apparent $G$-band magnitudes and parallaxes for each star remaining after the initial cut (black circles), $Gaia$ EDR3/APASS DR9 crossmatch (blue squares), and final sample (red triangles). These selections are applied sequentially, so the black sample also includes the red and blue points, and the blue sample includes the red points. The upper group of stars contains main sequence stars that are cut out by the series of selections.}
\label{fig:DataSel}
\end{figure}

\begin{figure}[ht!]
\begin{center}
    \includegraphics[angle=0,width=6in]{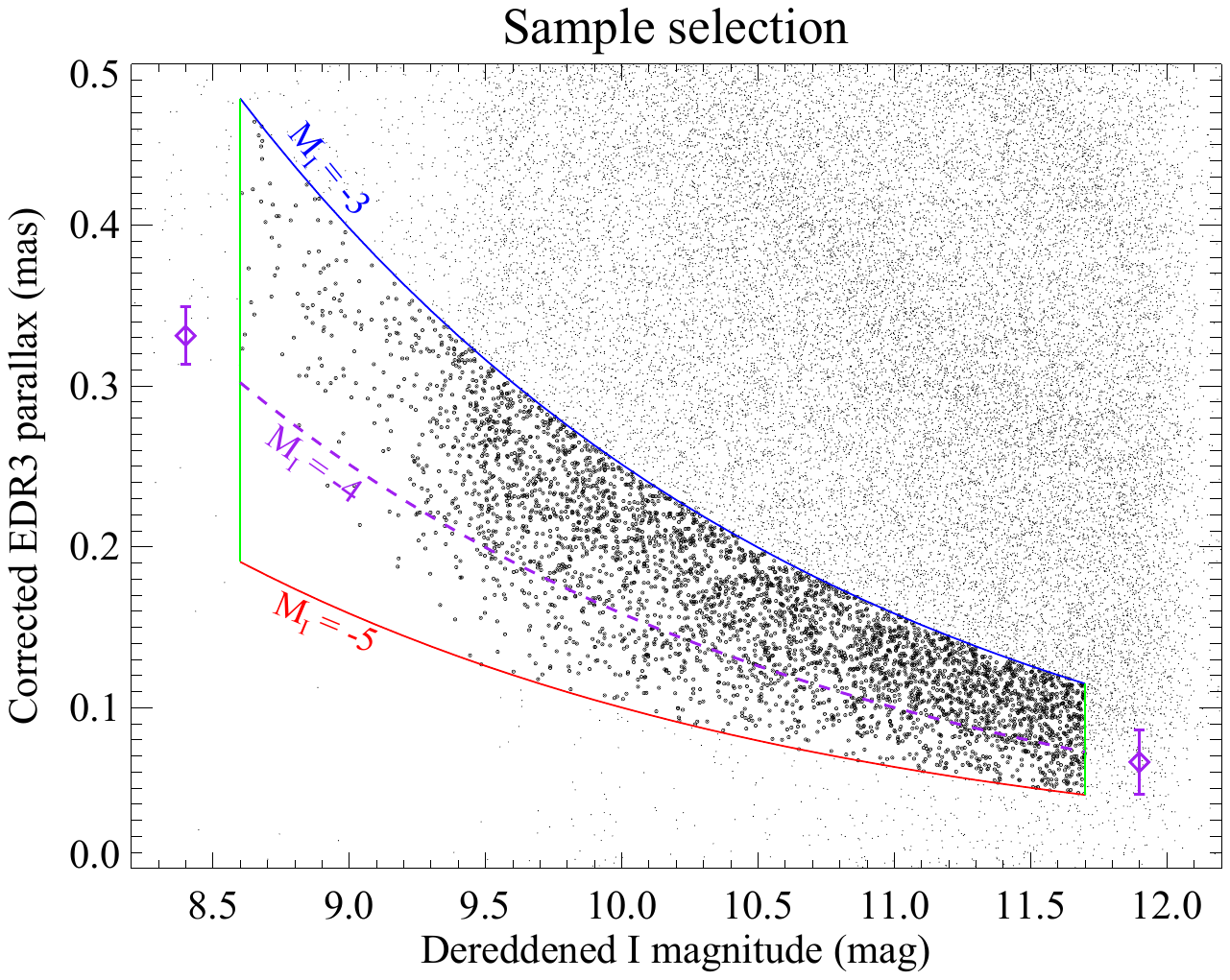}
\end{center}
\caption 
{Final sample after the parallax cuts in the $I$-band.  Parallax values include the \cite{Lindegren_2021AA...649A...4L} correction as well as the additional correction described in \S~\ref{sec:DataSelection}.  Stars within the selection boundaries are shown by heavier points.  The apparent magnitude selection ($8.6 < I < 11.7 $) is driven at the bright end by the saturation limit for APASS magnitudes, and at the faint end by the incompleteness of the sample approaching $ I \sim 12 $.  The parallax selection is based on the apparent magnitude in $ I $, and corresponds to the values derived from a nominal absolute magnitude of $ M_I = - 5 $~mag (red line) and $ -3 $~mag (blue line).  The error bars on the purple diamonds show typical (median) EDR3 parallax errors at those magnitudes; note that parallax errors change very little over this magnitude range.  The dashed purple line at $ M_I = 4.0 $~mag marks a representative location of the TRGB.  At bright magnitudes, corresponding to more nearby stars, the drop in density around $ -4.0 $ is apparent; at fainter magnitudes, covering the majority of the sample, the parallax uncertainties blur this transition.  These identical selections are used in the normalization integral of Eq.~\ref{eq:normalization_integral}.}
\label{fig:absrange}
\end{figure}

\begin{figure}[ht!]
\plotone{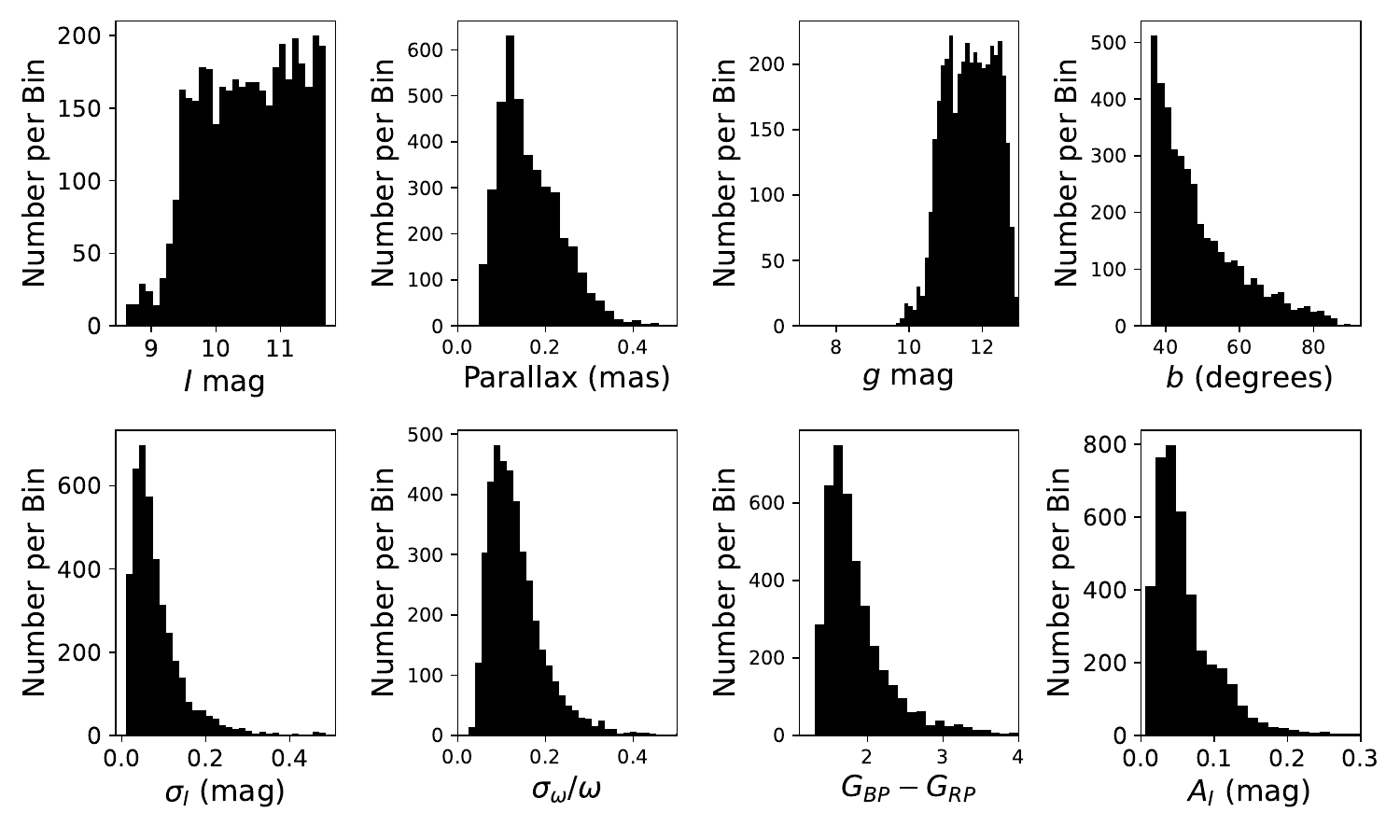}
\caption 
{Histograms for the final dataset after all selection cuts. In clockwise order from the top left: 1) $I$-band apparent magnitudes, 2) Parallax, 3) $Gaia$ $G$-band magnitudes, 4) Galactic latitude, 5) Extinction using the \cite{Schlafly_2011ApJ...737..103S} dust maps and NASA/IPAC extinction tool, 6) $Gaia$ $B_p - R_p$ color, 7) Fractional parallax error, 8) $I$-band apparent magnitude error. The abrupt cutoffs for 3, 4, and 6 are due to the initial $Gaia$ selection cuts.}
\label{fig:DataChar}
\end{figure}

After these selections, the final sample used in the likelihood calculation consists of 4,034 stars. We make these available in Table~\ref{tab:data}. The typical magnitude and parallax errors in this final sample are 0.084 $\pm$ 0.065~mag and 0.019 $\pm$ 0.008~mas,  respectively. In Fig. \ref{fig:DataSel} we plot the $ G $-band magnitudes and parallaxes of the stars remaining at three selection points, in Fig. \ref{fig:absrange} we plot the final sample in the context of the parallax cuts, and in Fig. \ref{fig:DataChar} we plot histograms for the $I$-band magnitude, $I$-band error, parallax, fractional parallax error, $G$-band magnitude, $ G_{BP} - G_{RP} $ color, galactic latitude, and extinction for the final sample.

\begin{deluxetable}{cccccccccc}
\tablecaption{Milky Way field Giants in Final Sample}
\label{tab:data}
\tablehead{\colhead{$Gaia$ ID} & \colhead{$G$ mag} & \colhead{$ G_{BP} - G_{RP} $} & \colhead{$I$ mag} & \colhead{$\sigma_I$} & \colhead{$\varpi^a$} & \colhead{$\sigma_{\varpi}$} & \colhead{sin$b$} & \colhead{$A_I$} & \colhead{$\sigma_{A(I)}$}}
\startdata
1092743175930914560 & 10.993 & 2.098 & 9.636 & 0.104 & 0.207 & 0.024 & 0.5969173669870012 & 0.083 & 0.002 \\
2847332664620076800 & 12.299 & 2.112 & 10.993 & 0.112 & 0.137 & 0.019 & 0.6303934482746393 & 0.089 & 0.002 \\
2837309379183101056 & 10.429 & 1.946 & 8.989 & 0.090 & 0.373 & 0.021 & 0.6088638645647745 & 0.260 & 0.020 \\
3488861793501000320 & 12.077 & 1.550 & 10.943 & 0.140 & 0.160 & 0.014 & 0.6157992207440517 & 0.106 & 0.002 \\
3489382480977057408 & 10.474 & 1.771 & 9.396 & 0.021 & 0.247 & 0.020 & 0.603055402691334 & 0.121 & 0.002 \\
6885457274187379584 & 12.475 & 1.471 & 11.597 & 0.163 & 0.103 & 0.014 & 0.6083110570577946 & 0.084 & 0.001 \\
6885690611170860800 & 12.429 & 1.436 & 11.477 & 0.032 & 0.072 & 0.013 & 0.6094298299352428 & 0.070 & 0.001 \\
6885710333660610048 & 12.461 & 1.439 & 11.520 & 0.051 & 0.092 & 0.013 & 0.604022444144464 & 0.065 & 0.002 \\
6885762075130437760 & 12.255 & 1.616 & 11.138 & 0.055 & 0.102 & 0.018 & 0.5968548945903711 & 0.082 & 0.003 \\
6891589869010926464 & 11.542 & 2.125 & 10.304 & 0.059 & 0.147 & 0.022 & 0.6366263135563972 & 0.074 & 0.004 \\
\enddata
\tablecomments{(a) \cite{Lindegren_2021AA...649A...4L} and \cite{Lindegren_2021_talk} offsets applied. Only a portion of this table is shown here to demonstrate its form and content. A machine-readable version of the full table is available.}
\end{deluxetable}

\section{Maximum Likelihood}

\label{sec:ML}

\subsection{Approach}
\label{ss:ML_details}

Our application of the ML approach involves a population of red giant stars in the Milky Way that can be described with luminosity function and density distribution models but have unknown model parameters. Our goal is to find the set of model parameters that maximizes the likelihood of observing the parallaxes and magnitudes of our sample of stars.

The likelihood of observing the parallax and magnitude of a single star for a set of fixed model parameters (which will be optimized later) can be expressed as a double integral of the form: 

\begin{equation}
\label{eq:likelihood}
    P(\varpi, m_I | \Lambda) = \int^{M_2}_{M_1} p_I(M_I) dM_I \int^{\infty}_{0} p_D(D) dD * p_{obs}(\varpi, m_I| D, M_I, \sigma_\varpi, \sigma_m)
\end{equation}

\noindent where $\varpi$ is the measured parallax, $m_I$ is the observed apparent magnitude in the Johnson-Cousins I-band corrected for extinction, $\sigma_\varpi$ is the error in the measured parallax, $\sigma_m$ is the error in the observed apparent magnitude, $\Lambda$ denotes the model parameters, $M_I$ is the true I-band absolute magnitude, $D$ is the true distance from the Sun, $p_I(M_I)$ is the luminosity function model, $p_D(D)$ is the density distribution model, and $p_{obs}$ is the joint probability of observing a star having a parallax $ \varpi $ and magnitude $m_I$ given measurement errors $\sigma_\varpi $ and $\sigma_m$. Assuming a Gaussian error model, $p_{obs}$ can be expressed as 

\begin{equation}
\label{eq:prob_guassian}
    p_{obs}(\varpi, m_I| D, M_I, \sigma_{\varpi}, \sigma_m) = C\;  \exp(-(\varpi - 1/D)^2 / (2\sigma^2_\varpi))\;\exp(-(m_i - M_I - dmod(D))^2 / (2\sigma_m^2))
\end{equation}

\noindent where C is the normalization and $dmod(D)$ is the distance modulus corresponding to the true distance $D$: 

\begin{equation}
    dmod(D) = 10 + 5\log_{10}(D)
\end{equation}

\noindent if $D$ is given in kpc. We note that here $p_{obs}$ only depends on measured properties of the star and is independent of the model parameters.

For our luminosity function model, $p_I$, we adopt a broken power law model following \cite{Makarov_2006AJ....132.2729M}:

\begin{equation}
\label{eq:LF}
\psi = \begin{cases}
10^{\alpha(M - M_{TRGB}) + \beta} & \quad M - M_{TRGB} \ge 0 \\
10^{\gamma(M - M_{TRGB})}         & \quad M - M_{TRGB} < 0
\end{cases}
\end{equation}

\noindent where $\alpha$ is the red giant branch (RGB) slope, $\beta$ is the strength of the TRGB break

\begin{equation}
    \beta = \log_{10}  \left(\frac{N - \epsilon}{N + \epsilon} \right)
\end{equation}

\noindent where $N - \epsilon$ and $N + \epsilon$ are the number of stars immediately brighter and fainter than the break, respectively, $\gamma$ is the AGB slope, $m$ is the magnitude, and $M_{TRGB}$ is the magnitude at the TRGB break.

For our density distribution model, $p_D$, we find that the distribution of distances from $Gaia$ EDR3 is best described by a density distribution model of the form 

\begin{equation}
\label{eq:DensityModel}
    p_D(D) = D^2 \exp(-D \cdot \sin(b)/h)
\end{equation}

\noindent where $h$ is the scale height and $b$ is the Galactic latitude for a given star.  We have also attempted to use multiple exponential components, but we find that one component is sufficient to match the observed sample; additional components are unconstrained.  We emphasize that this simple density model is optimized for the specific sample of stars under consideration; other samples, e.g., covering a broader range of distances / parallaxes, may require different density models.

Finally, it is critical that the likelihood computed from the probability in Equation~\ref{eq:likelihood} be normalized to the full extent of the sample.  Here the selection in measured parallax shown in Figure~\ref{fig:absrange} comes into play.  In essence, the maximum likelihood method maximizes the probability of the model parameters given where the observed quantities are distributed {\it within the selection boundaries}.  Thus the likelihood $ L (\varpi, m_I | \Lambda) $ for each star is expressed as the integral:
\begin{equation}
    L (\varpi, m_I | \Lambda) = P(\varpi, m_I | \Lambda) / N (\Lambda)
    \label{eq:normalized_likelihood}
\end{equation}
where the normalization $ N (\Lambda) $ is expressed by:
\begin{equation}
    N(\Lambda) = \int_{m1}^{m2} dm_I' \int_{{\varpi_{\rm min}(m_I')}}^{{\varpi_{\rm max}(m_I')}} d\varpi' \, \cdot P(\varpi', m_I' | \Lambda) \label{eq:normalization_integral}
\end{equation}
\noindent where $ m1, m2 $ are the apparent magnitude limits for the sample, and $ \varpi_{\rm min}, \varpi_{\rm max} $ are the minimum and maximum parallaxes accepted for a given value of the observed magnitude $ m_I $ (red and blue lines in Fig.~\ref{fig:absrange}).  We note that while the normalization integral is independent of the measured parallax and magnitude, it {\it does} depend on their measurement uncertainties, via the term $ p_{obs} $ in Equation~\ref{eq:likelihood}.  Therefore a separate calculation of the normalization integral is necessary for each combination of uncertainties in magnitude and parallax---effectively for each star in the sample.

Equation \ref{eq:normalized_likelihood} gives the likelihood $ L (\varpi, m_I | \Lambda) $ of observing a single star given a set of model parameters. To find the total likelihood of observing all the stars in the sample with a given set of model parameters, we would in principle take the product of all the individual likelihoods. In practice, we take the sum of the logarithm of the individual likelihoods to obtain a log likelihood $\mathcal{L}$ defined as:

\begin{equation}
    \mathcal{L}(\Lambda | \{\Theta_i\}) = \sum_i \log L(\Theta_i | \Lambda)
\label{eq:Total_Lik}
\end{equation}

\noindent where $\Theta_i$ denotes the observed parameters for star $ i $, and $ \{ \Theta_i \} $ the ensemble of such parameters for the full sample..

\subsection{Optimization: the shape of the luminosity function}
\label{sec:Optimization}

To find the set of parameters that maximizes the likelihood (or, equivalently, the log likelihood) of observing our sample of stars, we multiply the total log likelihood from Equation \ref{eq:Total_Lik} by $-1$ and minimize with respect to the model parameters using the Broyden–Fletcher–Goldfarb–Shanno (BFGS) method \citep{Broyden_10.1093/imamat/6.1.76, Fletcher_10.1093/comjnl/13.3.317, Goldfarb_10.2307/2004873, Shanno_10.2307/2004840}. We find that allowing any of the LF parameters to vary freely can cause the optimization to converge onto unphysical parameters. For instance, allowing all of the LF parameters to vary freely will cause $M_{TRGB}$ to converge at the faint end of the photometric cut together with a negative RGB slope. Because the fainter photometric cut creates an artificial discontinuity that is stronger than the actual TRGB break, the ML algorithm prefers this photometric cut over the actual TRGB break. \cite{Makarov_2006AJ....132.2729M} encountered a similar issue and attributed their case to the TRGB being too close to the photometric limit of their \textit{HST} images. While  \cite{Makarov_2006AJ....132.2729M} find that they can avoid this issue for galaxies with their TRGB more than $\sim$ 1~mag from the faint photometric cutoff, we are unable to avoid it here due to the two-dimensional nature of our optimization, which gives the parameters and absolute magnitudes more freedom to vary.

To avoid converging onto unphysical parameters, we constrain the optimization by holding the shape parameters for the luminosity function to values consistent with those found in external galaxies.  Specifically, we allow two parameters to vary: the absolute magnitude of the break, $M_{TRGB}$, and the scale height of the Galaxy, $h$, while we hold the parameters $\alpha $, $ \beta $, and $\gamma$ in Equation~\ref{eq:LF} fixed during the optimization. \cite{Mendez_2002AJ....124..213M} found that an assumed slope of $ \alpha = 0.3 $ can be used to obtain an accurate estimate of the TRGB. In addition, \cite{Makarov_2006AJ....132.2729M} use two alternative fitting constraints where they either fix or impose priors on both $\alpha$ and $\gamma$. We adopt the first of the two alternative constraints from \cite{Makarov_2006AJ....132.2729M} and fix both $\alpha$ and $\gamma$ to 0.3, as we find that our data do not constrain these parameters significantly.  Doing so allows us to successfully avoid converging onto the artificial maximum located at the cut.

The $\beta$ parameter measures the logarithmic strength of the break, and it depends on the relative number of AGB and RGB stars above and below $M_{TRGB}$.  In consequence, $ \beta $ can and does vary significantly from galaxy to galaxy, depending on each system's stellar properties, such as their star formation history.  This parameter is also more difficult to constrain than in other studies because of the two-dimensional nature of our optimization; we find that for our sample, $ \beta $ correlates with $M_{TRGB}$, creating a quasi-degeneracy in the optimization (larger values of $ \beta $ correspond to brighter $ M_{TRGB} $). A heuristic explanation for this behavior is that for larger $ \beta $, the probability of finding stars brighter than the break is smaller; thus $ M_{TRGB} $ is pushed brighter to better accommodate stars on the bright side of the jump.  

Formally, the fit for $ \beta $, with $ \alpha $ and $ \gamma $ fixed, converges to a value of $ M_{TRGB} = -4.00 $~mag; however, the fit for $ \beta $ is poorly constrained, with a nominal value of $ \beta = 0.82 $.  To verify that this fit is reasonable, we have looked for the value of $\beta$ in galaxies that are similar to the Milky Way in structure and composition and have well defined TRGB breaks. We select galaxies that have photometry available on the Extragalactic Distance Database \citep[EDD;][]{Anand_EDD_2021AJ....162...80A} website\footnote{http://edd.ifa.hawaii.edu/} and are less than 15 Mpc away, are Sab to Sbc type, have an inclination of greater than 70\textdegree, have more than 2000 stars in the $\pm 1$~mag range around the EDD TRGB, and have a $B_T$ luminosity within 2~mag of -19.7 mag. This selection leaves NGC~253, NGC~891, NGC~4258, and NGC~4565. We note that while NGC~7090 also satisfies these criteria, we exclude it because its photometry shows two discontinuities within 1 mag, which can bias the LF parameters.

We apply one-dimensional ML fits to these galaxies using the same LF model in Equation \ref{eq:LF} to find their LF parameters. While EDD have performed ML fits on these galaxies in the past, they have not yet published their ML parameters.  We retrieved the photometry for these galaxies from the EDD website and carried out a one-dimensional ML optimization using the BFGS method; our methodology differs somewhat from that of EDD. We also compare the value of the TRGB when allowing $\beta$ to vary freely to the TRGB with these constraints in Section \ref{subsec:OptimizationResults}.

The results of the one-dimensional ML fits are reported in Table~\ref{tab:Galaxy_1D_ML_Fits}.  The values of $ \beta $ resulting from this process lie between $ 0.067 $ for NGC~4565 and 0.321 for NGC~891, significantly smaller than the nominal fit for the MW sample. We tentatively adopt these as external constraints for $ \beta $ in the MW.  To investigate possible variations of $\beta$ across different fields of a given galaxy, we apply a ML fit to the second NGC 4258 field available on EDD (GO-16198) and find a $\beta$ of 0.240, which is 0.012 larger than the NGC 4258 $\beta$ we found for the first field (GO-9477) and listed in Table \ref{tab:Galaxy_1D_ML_Fits}; this suggests that such variations are much smaller than the range in $\beta$ across different galaxies.

\subsection{Optimization results: the magnitude of the TRGB}
\label{subsec:OptimizationResults}

As discussed in \S~\ref{sec:Optimization}, we carry out the two-dimensional maximum likelihood optimization to solve for $ M_{TRGB} $ and $ h $ for fixed values of the luminosity function shape parameters $ \alpha $, $ \beta $, and $ \gamma $. We adopt $ \alpha = \gamma = 0.3 $, and we consider two bracketing values for $ \beta $, namely $ \beta = 0.067 $ and $ \beta = 0.321 $; in addition, we also run an optimization with $ \beta $ as a free parameter.  For each set of values, we conduct the maximization of the quantity $ \cal L $, defined in Equation~\ref{eq:Total_Lik}, on the Maryland Advanced Research Computing Center (MARCC) computer system using the BFGS method.  The statistical uncertainties in the parameters are computed on the basis of the Hessian matrix produced by the BFGS solver.

\begin{figure}[ht!]
\hbox to \hsize{\hss\includegraphics[angle=0,width=3.5in]{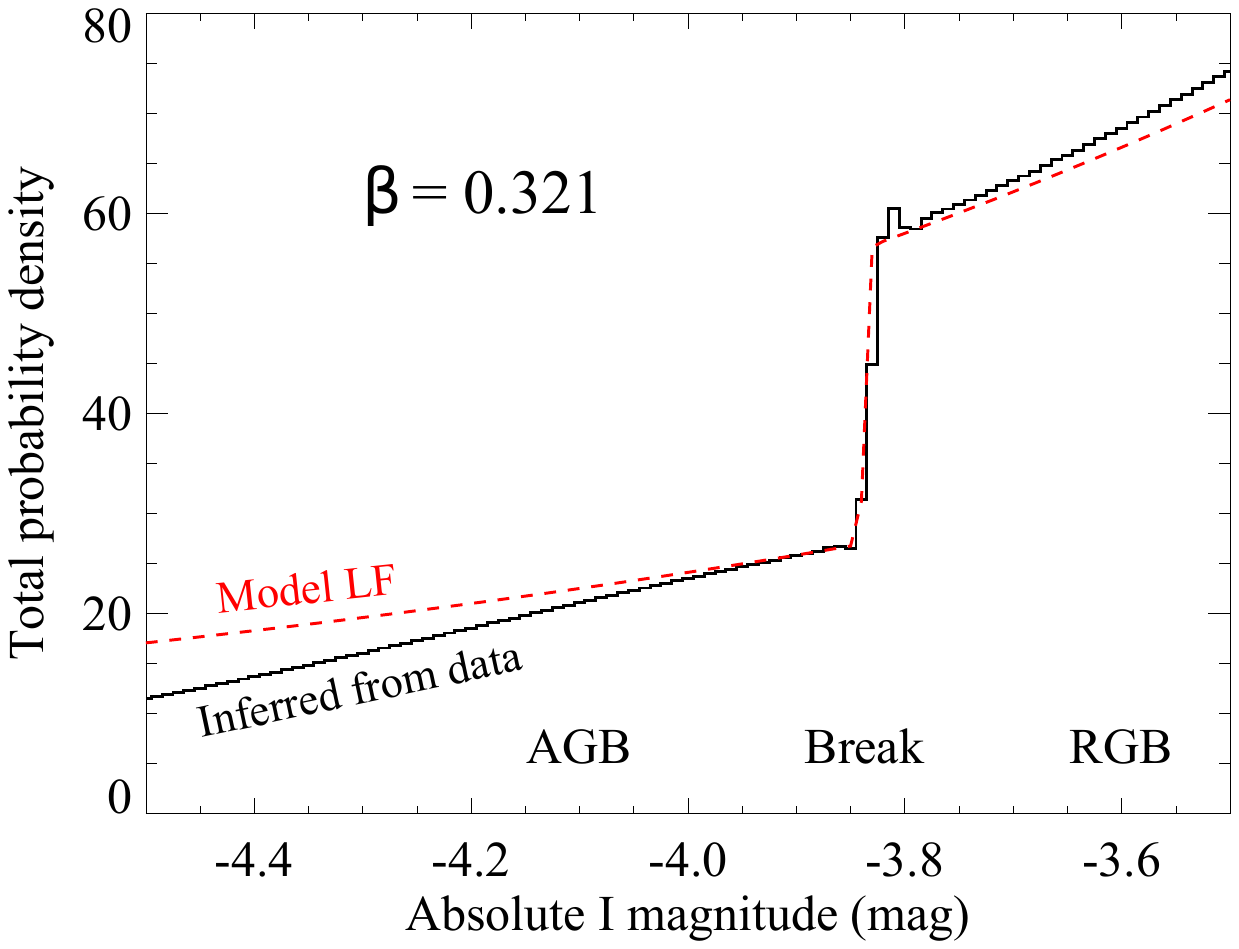}
\hss
\includegraphics[angle=0,width=3.5in]{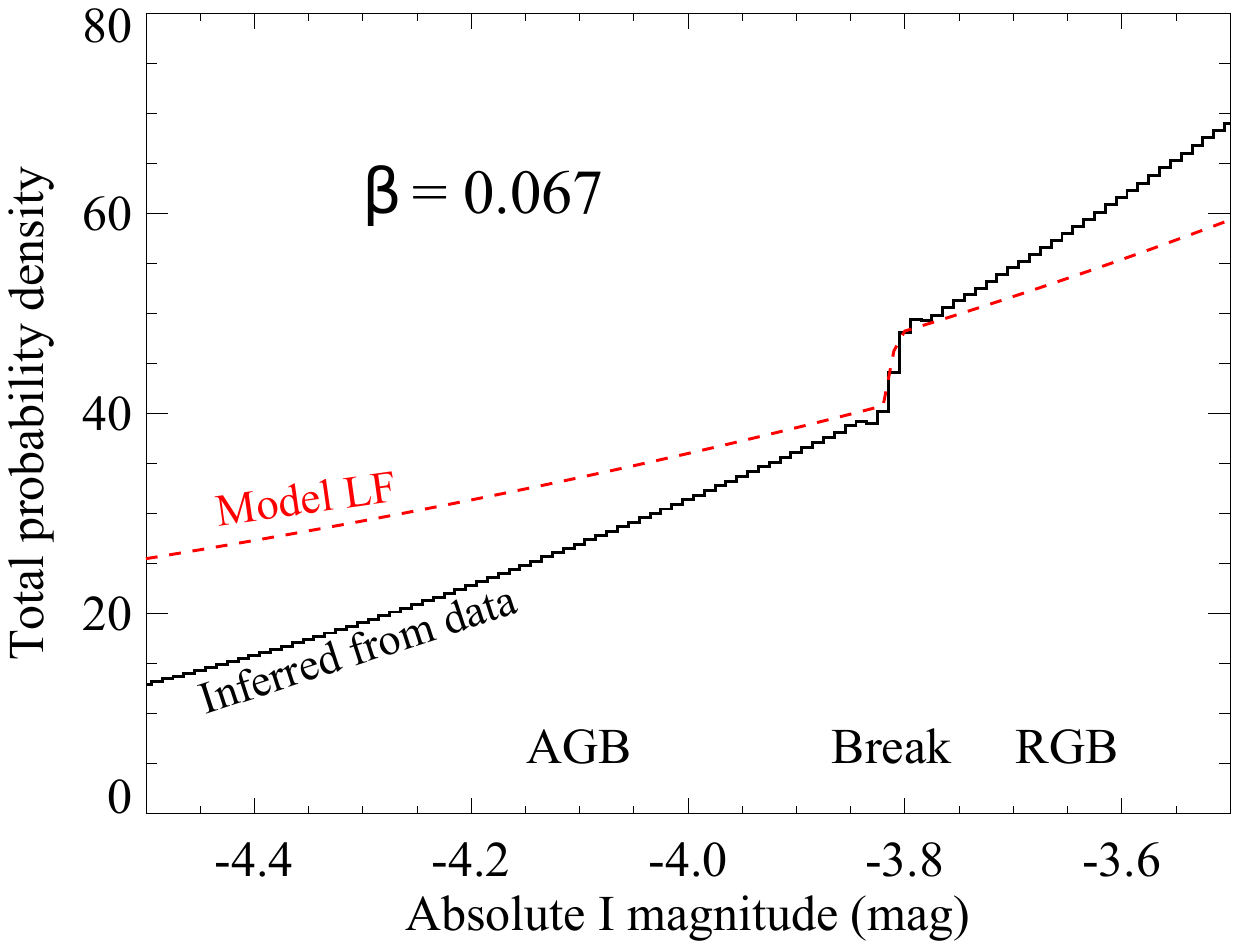}\hss}
\caption {Probability distribution of absolute magnitude for the best-fit model for the two values which bracket the expected range of $ \beta $ listed in the text.  The black histograms show the absolute magnitude probability distribution, marginalized in distance and summed over all the stars in the sample, i.e., the most likely distribution of their absolute magnitudes.  The red lines show the corresponding model or idealized luminosity functions.  The small blips on either side of the break are a result of the approximation used in drawing this figure, and can be ignored.  The results shown here do not yet account for the empirical relationship between the color and luminosity of the TRGB, which we determine to be $\sim$ 0.05 mag between the observed and a fiducial, blue TRGB.  See \S~\ref{subsec:OptimizationResults} for further details.}
\label{fig:PostDist}
\end{figure}

Using $\beta$ values of 0.321 and 0.067, we find $M_{TRGB}$ values of $-3.84$ $\pm$ 0.05~mag and $-3.81$ $\pm$ 0.07~mag, and $h$ values of 2.30 $\pm$ 0.01~kpc and {2.13 $\pm$ 0.02}~kpc, respectively.  Figure~\ref{fig:PostDist} shows the inferred probability distribution of absolute magnitude for these best-fit models.  In each panel, the black histograms show the probability distribution of the absolute magnitude for the measured apparent magnitude and parallax of each star, given their measurement errors, position on the sky, and the luminosity function for that model; the probability is marginalized over distance and summed over all stars in the sample.  The red lines show the corresponding model luminosity functions.  Because of the measurement uncertainties, especially in apparent magnitude, the size of the break is driven primarily by the model luminosity function itself; for a star with apparent magnitude and parallax consistent with the break, the relative probability of being just above or just below the break is primarily driven by the luminosity function.  Smaller measurement errors will greatly increase the ability of the data to constrain the size of the break.  However, other features, such as the slope of the luminosity probability function away from the break, are less sensitive to measurement errors, and depend also on the distribution of measured quantities; hence the probability distribution can deviate in slope from the underlying assumed luminosity function.  Note that close to the edges of the adopted range in absolute magnitude, selection effects---accounted for in the maximum likelihood normalization---can affect on the reconstructed probability distribution.

The fit with an unconstrained value of $ \beta $ converges to a TRGB of $ M_{TRGB} =$ -4.00, $\beta$  of 0.82, and $h$ = 2.51 ~kpc.

For the purpose of the current analysis, we consider the difference between the two values of $ M_{TRGB} $ to be a {\it systematic} uncertainty.  Therefore we adopt a value   $ M_{TRGB} = -3.91$~mag, the midpoint of the two extreme cases; to account for the variation in $M_{TRGB}$ from the range of $\beta$, we add a systematic error of 0.09~mag, which is the distance between the endpoint $M_{TRGB}$ values and the midpoint and encapsulates the statistical errors which are all smaller than the systematic differences that set this error. We list the error budget and anticipated future improvements in Table \ref{tab:Error_Budget} and expect this measurement to improve with a future replacement of APASS photomtery with $Gaia$ DR3 photomtery.

In order to compare our inferred value of $ M_{TRGB} $ with other results, we need to take into account the color of our sample and correct our TRGB to a fiducial color.  The mean $V-I$ color of our sample is 1.72~mag. Because \cite{Jang_2017ApJ...835...28J} find that colors bluer than $V-I$ = 1.9~mag do not need a color correction, we do not apply a color correction to our TRGB measurement. 

\begin{deluxetable}{ccc}
\tablecaption{TRGB Zero-point Error Budget and Estimated Future Improvements}
\label{tab:Error_Budget}
\tablehead{\colhead{Source of Uncertainty} &\colhead{$\sigma_{stat}$} & \colhead{$\sigma_{sys}$}}
\startdata
Maximum Likelihood & 0.05 & \\
$\beta$ &  & 0.09 \\
\hline
Total & 0.05 & 0.09  \\
\hline
\multicolumn{3}{c}{Factor Improvement with Future Data Releases} \\
\hline
\textit{Gaia DR3} & $ \sim 3.1 $ & \\
\textit{Gaia DR4} & $ \sim 3.6 $ & 
\enddata
\tablecomments{Error budget and sensitivity to measurement errors for the TRGB measured in this study. The last two lines are the expected improvements in error with \textit{Gaia} DR3 and DR4.}
\end{deluxetable}


\begin{deluxetable}{ccc}
\tablecaption{Milky Way TRGB derived from Reference Galaxy $\beta$}
\label{tab:Galaxy_1D_ML_Fits}
\tablehead{\colhead{Galaxy} &\colhead{$\beta$} &   \colhead{$M^{TRGB, a}_{MW, I}$}}
\startdata
NGC 4258 & 0.228 (0.024)& -3.82 (0.04) \\
NGC 253 &  0.181 (0.013)  & -3.81 (0.04)  \\
NGC 4565 & 0.067 (0.008) & -3.81 (0.07) \\
NGC 891 & 0.321 (0.029) & -3.84 (0.05)
\enddata
\end{deluxetable}

\begin{deluxetable}{lllllll}
\tablecaption{Recent Calibrations of the TRGB in the $I$-band (since 2017)}
\label{tab:TRGBComparison}
\tablehead{\colhead{Study} & \colhead{$I^{TRGB}$} & \colhead{Field} & \colhead{Distance} & \colhead{Extinction ($A_I$)} & \colhead{$M^{TRGB}_I$} & \colhead{$\sigma$}}
\startdata
\multicolumn{6}{c}{NGC 4258, $\mu$ = 29.397 $\pm$ 0.024 (stat) $\pm$ 0.022 (sys) mag from \cite{Reid_2019ApJ...886L..27R}} \\
\hline
\cite{Freedman_Tensions_2021ApJ...919...16F,Jang_2021ApJ...906..125J} & 25.372 (0.015) & HST 9477, 10399& & 0.025 & $−4.043^a$ & 0.056  \\
\cite{Anand_EDDvsCCHP_2021arXiv210800007A} & 25.427 (0.027) & HST 16198&  & 0.025 & $−4.003^{a,i}$ & 0.042  \\
\cite{Anand_EDD_2021AJ....162...80A} & 25.43 (0.03) &  HST 9477 & &0.025 & $−4.00^{a}$ & 0.04  \\
\hline
\multicolumn{6}{c}{LMC, $\mu$ = 18.477 $\pm$ 0.004 (stat) $\pm$ 0.026 (sys) mag from \cite{Pietrzynski_2019Natur.567..200P}} \\
\hline
\cite{Freedman_Tensions_2021ApJ...919...16F,Hoyt_2021arXiv210613337H} & 14.56 (0.015)$^k$ && &  0.12$^e$ & $−4.038$ & 0.036 \\
\cite{Yuan_2019ApJ...886...61Y} & 14.607 (0.023) & &&   0.10$^d$ $\pm 0.03$  & $-3.963$ & 0.046 \\
\cite{Jang_2017ApJ...835...28J} & 14.624 (0.020)& &&   0.10$^d$ $\pm 0.07$  & $-3.953^b$ & 0.102 \\
\cite{Gorski_2020ApJ...889..179G} & 14.635 (0.02)  &  && 0.12$^e$ & $−3.96$ & 0.04   \\
\hline
\multicolumn{6}{c}{SMC, $\mu$ = 18.977 $\pm$ 0.016 (stat) $\pm$ 0.028 (sys) mag from \cite{Graczyk_2020ApJ...904...13G}} \\
\hline
\cite{Freedman_Tensions_2021ApJ...919...16F,Hoyt_2021arXiv210613337H} & 14.927 $(0.022)^d$ & & (e) & $-4.050$ & 0.050 \\ 
\hline
\multicolumn{6}{c}{MW, Omega Centauri Globular Cluster} \\
\hline
\cite{Freedman_Tensions_2021ApJ...919...16F,Cerny_2020arXiv201209701C} & (f) && 13.678 $(0.11)^g$ & (h) & $-4.056$ & 0.102  \\
\cite{Soltis_2021ApJ...908L...5S} & 9.84 (0.04)& & 13.595 (0.047) & 0.215 $\pm 0.011$ & $−3.97$ & 0.06 \\
\cite{Appelaniz_2021AA...649A..13M} & " && 13.60 (0.11) & " & $-3.98$ & 0.12 \\
\cite{Vasiliev_2021MNRAS.505.5978V} & " && 13.57 (0.10) & " & $-3.95$ & 0.11 \\
\cite{Capozzi_2020PhRvD.102h3007C} & " & &13.597 (0.021) & " & $−3.96$ & 0.05 \\\hline
\multicolumn{6}{c}{MW, 47 Tuc two DEBs tied to 12 Globular Clusters} \\
\hline
\cite{Freedman_2020ApJ...891...57F} & (f)& & 13.27 $(0.07)^j$ &  (f) & $−4.056$ & 0.096 \\
\cite{Vasiliev_2021MNRAS.505.5978V} & (f)& & 13.17 $(0.08)$ &  (f) & $−3.96$ & 0.10 \\
\cite{Appelaniz_2021AA...649A..13M} & (f)& & 13.23 $(0.09)$ &  (f) & $−4.02$ & 0.11 \\
\cite{Baumgardt_2019MNRAS.482.5138B} & (f) & & 13.24 $(0.02)$ &  (f) & $−4.03$ & 0.07 \\
\hline
This Study &&& &&${-3.91}$ & 0.05 (stat) \\
&&&&&& 0.09 (sys) \\
\hline
\enddata
\tablecomments{(a) Zero-points in $F814W$ transformed to $I$ by adding 0.0068~mag from \cite{Freedman_2019ApJ...882...34F}, (b) The original $M^{TRGB}_I$ = $-3.970$ mag used a distance of $\mu$ = 18.494 $\pm$ 0.008 (stat) $\pm$ 0.048 (sys) mag from \cite{Pietrzynski_2013Natur.495...76P},  (d) Extinction corrected $I^{TRGB}_0$ using \cite{Haschke_2011AJ....141..158H}.  (e) Extinction corrections were performed using location using maps from \cite{Skowron_2021ApJS..252...23S}. No statistic for these extinction corrections was provided, (f) Multiple globular clusters were used in this study anchored to one, (g) \cite{Thompson_2001AJ....121.3089T}, (h) Extinction corrections for multiple globular clusters were performed with \cite{Harris_2010arXiv1012.3224H}, (i) includes color transformation applied by \citet{Anand_EDDvsCCHP_2021arXiv210800007A}, to facilitate comparisons, (j) \cite{Thompson_2020MNRAS.492.4254T}, (k) Calculated using the extinction corrected TRGB of 14.44~mag for Rank 1 $\&$ 2 fields from Fig. 15 and in the text in \cite{Hoyt_2021arXiv210613337H} and subtracting $A_I$ = 0.12~mag using the mean extinction for Rank 1 $\&$ 2 fields of E(V-I) = 0.10 from Figure 3, also in \cite{Hoyt_2021arXiv210613337H}.}
\end{deluxetable}

\begin{figure}[ht!]
\plotone{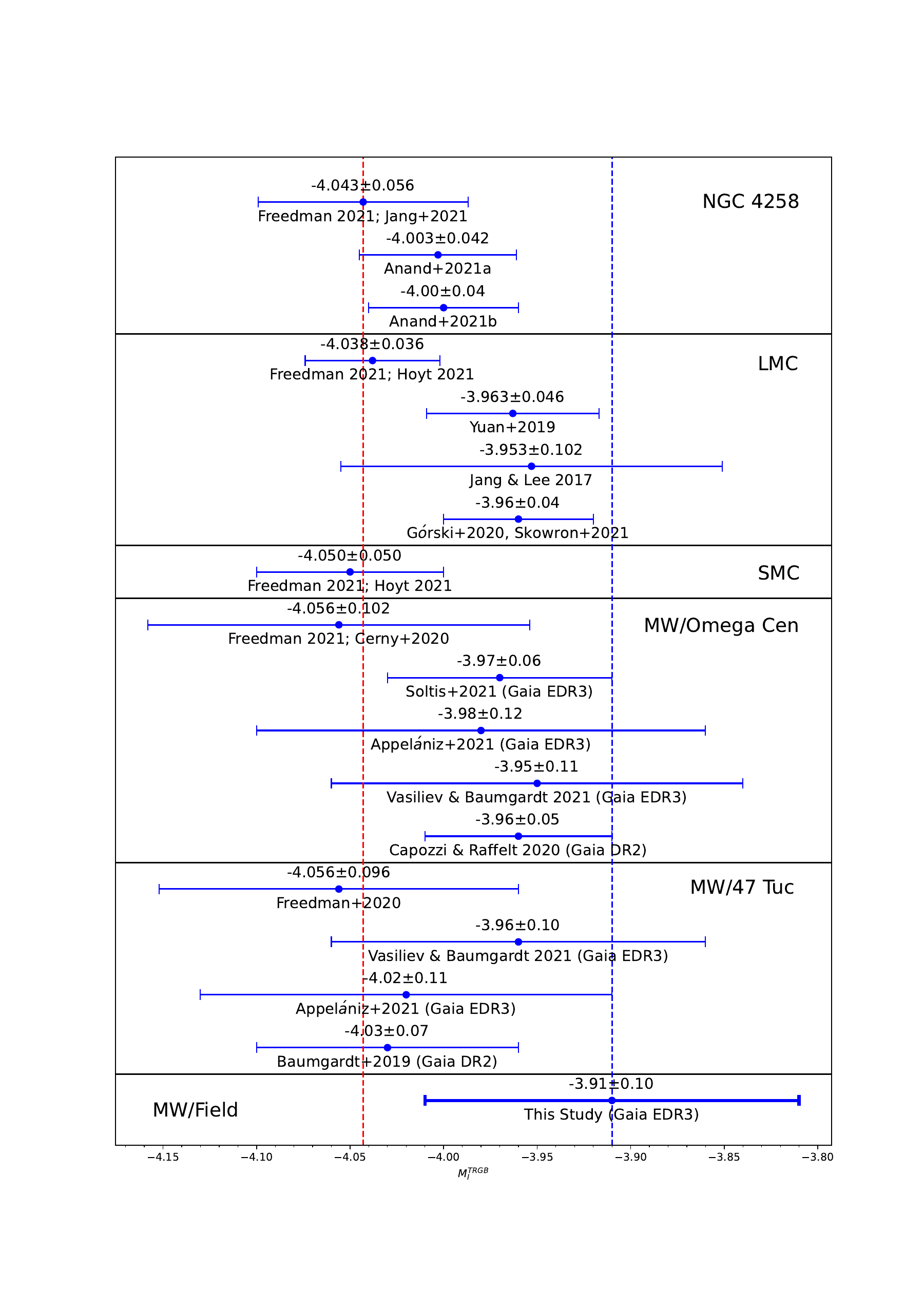}
\caption 
{The plot shows a comparison TRGB zero-point calibrations since 2017 separated by galaxy. The right blue dashed vertical line represents the midpoint TRGB value of $ -3.91 $~mag found here. The left red dashed vertical line shows the weighted average of the 4 measures from \cite{Freedman_Tensions_2021ApJ...919...16F} of $M^{TRGB}_{F814W} = -4.043 \pm 0.025$~mag.  Changing their value by 0.05~mag would change the derived value of $ H_0 $ from 69.8 to 71.4~km s$\textsuperscript{−1}$ Mpc$\textsuperscript{−1}$.}
\label{fig:TRGB_Comp}
\end{figure}

\clearpage
\section{Discussion}
\label{sec:Disc}

We have developed an independent TRGB zero-point calibration method that uses a two-dimensional ML approach applied to a sample of high-latitude field giants in the Milky Way, with parallaxes from $Gaia$ EDR3 and photometry from APASS.  We find that the value of the TRGB zero-point depends at the $\sim$ 0.05 mag level on the value of the $ \beta $ parameter in the luminosity function form proposed by \cite{Makarov_2006AJ....132.2729M}, which characterizes the magnitude of the LF discontinuity at the TRGB break.  Further, we find that the currently available data have magnitude and parallax uncertainties that are insufficient to significantly constrain the value of $ \beta $ for our sample of Milky Way giants.  Using a range of values for $ \beta $ from galaxies similar to the Milky Way with good TRGB determinations, we find values of $ M_I^{TRGB} $ ranging from $ -3.88$ $\pm$ $0.07 $~mag to $ -3.84$ $\pm$ $0.05$~mag. Allowing $\beta$ to vary freely results in $  M_I^{TRGB} = -4.00$. These values are broadly consistent with other recent calibrations of the TRGB zero-point and offer a potentially interesting level of statistical precision; see Table~\ref{tab:TRGBComparison} and Fig.~\ref{fig:TRGB_Comp}.  However, the uncertainty in the value of $ \beta $ needs to be substantially reduced in order for this approach to achieve its statistical potential.  Additional constraints on $ \beta $ can in principle be obtained via population synthesis studies for the Milky Way and similar galaxies; such considerations go beyond the scope of the present work. 


To account for the parallax offset caused by thermal oscillations of $Gaia$, we considered the offsets calculated in \cite{Lindegren_2021AA...649A...4L}. However, these offsets are sparsely calibrated for some of the apparent magnitude range considered here ($ 8 < G < 13 $), and other studies  \citep[collected in][]{Lindegren_2021_talk} find that an additional offset may be needed in this magnitude range (Fig.~\ref{fig:PaxOffset}).  
Therefore we use the additional offset given in Equation~\ref{eq:delta_offset}.
Without this additional offset, the TRGB for $\beta = 0.321$ and 0.067 would be $-3.87 \pm 0.02 $~mag and $-3.85 \pm 0.07 $~mag, respectively. 
It will be important in forthcoming $Gaia$ data releases to improve the understanding and measurement of the parallax offset, and especially to reduce its dependence on magnitude, color, and position of the source.  

\subsection{Recent TRGB Luminosity Calibrations}

Several recent papers have reported values of the TRGB luminosity based on a variety of techniques.  Table~\ref{tab:TRGBComparison} reports recent measurements and includes all those provided since 2017 and that are not superceded by a revision of the same measure from the same authors.  These are grouped by the source used in the calibration: NGC~4258, the Large and Small Magellanic Clouds, globular clusters in the Milky Way, and finally field stars in the Milky Way (the present work).  A closer look at individual TRGB calibrations shows that they typically are based on three components: the measured apparent magnitude of the TRGB in the target source, $ I^{TRGB} $; the estimated extinction at the tip, $ A_I $; and the assumed geometric distance modulus $\mu_0$ of the host.  (This description is not applicable to the present study, in which TRGB sources are at different distances and apparent magnitudes.)  The absolute magnitude of the tip is then $M_I^{TRGB}=I^{TRGB}-A_I-\mu_0$.  In order to clarify the comparison of different results, we have collected in Table~\ref{tab:TRGBComparison} the relevant components of the calibration used or assumed in each study, whenever possible.  In one case, we have adjusted the published measurement to conform to a common distance assumption for the source; this applies to an LMC measurement \citep{Jang_2017ApJ...835...28J}, which used an earlier distance from \citet{Pietrzynski_2013Natur.495...76P} and has been corrected to the new distance estimate by \citet{Pietrzynski_2019Natur.567..200P}.  We also corrected the NGC~4258 measurements from $F814W$ to $ I $ by adding $ 0.0068 $~mag as recommended by \citet{Freedman_2019ApJ...882...34F}.  
On the other hand, all results shown for NGC 4258 use the same distance from \citet{Reid_2019ApJ...886L..27R} and extinction from \cite{Schlafly_2011ApJ...737..103S}, and all results for the MW tied to the globular cluster $\Omega$ Centauri use the same $I^{TRGB}$ from \cite{Bellazini_2001ApJ...556..635B} and extinction from \cite{Schlafly_2011ApJ...737..103S}.  See \cite{blakeslee:2021} Table 3 for a similar table.  The adjusted values are compared graphically in Figure~\ref{fig:TRGB_Comp}.

The reported values of $ M_I^{TRGB} $ in Table~\ref{tab:TRGBComparison} and Figure~\ref{fig:TRGB_Comp} cover a significant range, from $ -3.95 $ to $ -4.06 $~mag, $\sim$5$\%$ in $H_0$, with typical individual uncertainties ranging from $ 0.04 $ to $ 0.12 $~mag.  Our final result and estimated uncertainty cover most of the range of previously published values, and therefore do not provide yet a decisive constraint on the TRGB calibration; likely future improvements are discussed in \S~\ref{subsec:future}.


\subsection{TRGB concordance and discordance}

Despite the use of consistent distance and $ I^{TRGB} $ values, there appear to be noteworthy differences between TRGB calibrations.  In particular, we note that the four independent calibrations presented by  \cite{Freedman_Tensions_2021ApJ...919...16F} (hereafter F21), shown in Figure~\ref{fig:TRGB_Comp}, all appear to lie on the brighter side of the range of zero-points found in other studies, and are tightly grouped at $M_I^{TRGB} =-4.04$~mag (or $-4.05$~mag in $F814W$), shown as a vertical dashed line.  This calibration, together with TRGB measurements in several SN~Ia hosts, was used by F21 to obtain a Hubble Constant of $H_0 = 69.8 \pm 0.6$ (stat) $\pm 1.6$ (sys) km s$\textsuperscript{−1}$ Mpc$\textsuperscript{−1}$.

The other recent sources tend to be fainter and range from $-4.00$~mag to $-3.95$~mag (it is important to note that many of these group's measures are not independent from each other, as they may involve alternative measurements of a component from similar data) with the calibration from \cite{Anand_EDDvsCCHP_2021arXiv210800007A} of $-4.00$~mag used to obtain  a Hubble Constant of $H_0 = 71.5 \pm  1.9$ km s$\textsuperscript{−1}$ Mpc$\textsuperscript{−1}$.  This difference in luminosity primarily arises from differences in a {\it specific component} of the calibration at each location by $\sim$0.05--0.08~mag, which we highlight by adding additional literature measurements for the relevant component.  For NGC 4258 and LMC, the difference can be attributed to a brighter $I^{TRGB}$. For MW, this is due to further distance to $\Omega$ Centauri,  respectively, utilized by F21 relative to other sources. 

In the case of the LMC, the apparent tip ranges from $I^{TRGB}=14.64 \pm 0.02$~mag from the mean field in \cite{Gorski_2018AJ....156..278G} to $I^{TRGB}=14.56 \pm 0.015$~mag in F21, both using OGLE $I$-band photometry.  F21, based on  \cite{Hoyt_2021arXiv210613337H}, selects 8 of 25 LMC fields (``Rank 1 $\&$ 2''), whose TRGB measurement appears of the highest rank ``based on the observed quality of the TRGB detection''; these are brighter than the average field, although it is not clear how this selection process might conform to extragalactic measurements of average halo fields.\footnote{\cite{Freedman_2020ApJ...891...57F} and \cite{Freedman_2019ApJ...882...34F} which are superceded by F21 found the same LMC TRGB luminosity as F21 but with different component values that largely cancel a change, with the earlier $I^{TRGB}$ fainter by 0.04~mag and $A_I$ higher by 0.05~mag.}We note that the extinction applied to the four LMC entries in Table \ref{tab:TRGBComparison} differ by $\Delta A_I$ = 0.02~mag, and applying a common extinction would not be enough to account for the $\sim$ 0.07~mag difference across zero-point values suggesting that some of the difference also lies in different measurements of the location of the TRGB itself.

In NGC 4258, the difference between F21 (based on \cite{Jang_2021ApJ...906..125J}) and two different fields analyzed by  \cite{Anand_EDDvsCCHP_2021arXiv210800007A} and \cite{Anand_EDD_2021AJ....162...80A} arises from a 0.06~mag difference in $I^{TRGB}$ (or 0.04~mag after color standardization).   For the MW, F21, following \cite{Cerny_2020arXiv201209701C}, adopts the distance to $\Omega$ Centauri from one detached eclipsing binary system \citep{Thompson_2001AJ....121.3089T}. Recently, several studies have recalibrated the distance to $\Omega$ Centauri using $Gaia$ DR2 or EDR3; these estimates are all 0.08 to 0.10~mag closer than the older DEB distance, with uncertainties ranging from 0.02~mag \citep{Baumgardt_2019MNRAS.482.5138B} to 0.11~mag \citep{Appelaniz_2021AA...649A..13M}.  The present study based on field Giants and $Gaia$ parallaxes also yields a fainter tip and thus a higher Hubble Constant.  As we expect the precision and accuracy of $Gaia$ parallaxes and proper motions to improve with each data release, it appears that only $Gaia$ astrometry offers a path to improving the geometric foundation of the TRGB calibration in the Milky Way for the foreseeable future.  

As shown in Figure \ref{fig:TRGB_Comp}, the agreement of the groups of independent measurements is surprisingly good, but around two somewhat different means.  The four calibrations used by F21 have a total $\chi^2$ of $ 0.055 $ around their mean of $M_I=-4.043 \pm 0.025$~mag.  With $N=3$ degrees of freedom, the probability of such agreement (or better) is 0.3\% (1 in 300 or $\sim$ 3$\sigma$).   \cite{Freedman_2020ApJ...891...57F} give an additional calibration quite similar to the others in the globular cluster 47 Tuc of $-4.056$ $\pm$ 0.096 mag.  This is based on a DEB distance of $\mu=13.27 \pm 0.07$ mag, which is larger by 0.03 to 0.10 mag from those recently based on Gaia DR2 and EDR3 \cite{Baumgardt_2019MNRAS.482.5138B,Vasiliev_2021MNRAS.505.5978V,Appelaniz_2021AA...649A..13M,Soltis_2021ApJ...908L...5S}.  The other measurements for the same sources and not associated with the F21 group (see Fig.~\ref{fig:TRGB_Comp}) also show good internal agreement, though less surprising, with a $\chi^2=0.50$ (for $N=3$) for the four independent types in Figure 6 (top entry used) around a mean of $-3.982 \pm 0.024$~mag, an 8\% chance for such good agreement (1.7$\sigma$).  However, the two means are separated by 0.06 mag with a $2.4 \sigma$ ``tension'' based only on their respective formal uncertainties. Hypothetically adding 0.05~mag to the F21  value, to $ M_I^{TRGB} = -3.99 $~mag, would shift the inferred value of $ H_0 $ from 69.8 to to 71.4 km s$\textsuperscript{−1}$ Mpc$\textsuperscript{−1}$.  Comparisons of a small number of SN Ia hosts with distances measured by the TRGB or Cepheids unfortunately introduce additional variables and their uncertainties to the comparison and thus provide no additional insight to the TRGB calibration \citep{Riess:2022}.
Currently, the uncertainties in these zero-points make it difficult to determine whether a brighter or fainter zero-point is more accurate, but future TRGB zero-point calibrations may provide insight into this issue.

\subsection {Future Improvements}
\label{subsec:future}

Our current inability to constrain the shape of the luminosity function near the TRGB, and especially the parameter $ \beta $, which measures the strength of the break, is primarily due to the uncertainties in measured magnitude and parallax for stars in our sample.  We expect a substantial improvement in the quality of magnitude data with the upcoming release of $ Gaia $ DR3 (mid-2022), which will include wavelength-resolved photometry for most stars, allowing us to synthesize magnitudes in the desired passbands with expected uncertainty of about of $ 0.01 $~mag.  Simulations of such an improvement in magnitude accuracy suggest that DR3 results will enable a meaningful constraint in $ \beta $, and reduce the statistical error by a factor of $ \sim $ 3.  Further into the future, $Gaia$ Data Release 4 (and following) will yield significant improvements in parallax measurements.  Parallax errors of $\sim$ 0.01~mas, close to $ Gaia $ pre-launch expectations, will yield a further improvement of more than an additional 10$\%$ in the statistical errors of our TRGB determination.  At that time, Milky Way stars may well produce a calibration of $ M_I^{TRGB} $ with an accuracy better than 0.01~mag.

\section{Acknowledgments}

The authors would like to thank Dan Scolnic and Lucas M. Macri for their helpful comments. This research made use of the APASS database located at the AAVSO website and funded by the Robert Martin Ayers Sciences Fund, cross-match service provided by CDS, Strasbourg, NASA/IPAC Infrared Science Archive which is funded by the National Aeronautics and Space Administration and operated by the California Institute of Technology, and computational resources at MARCC.

\bibliography{1.TRGB}{}
\bibliographystyle{aasjournal}

\end{document}